\documentstyle[psfig]{cupconf}

\ifoldfss
\else
  \ifnfssone
    \newmathalphabet{\mathit}
      \addtoversion{normal}{\mathit}{cmr}{m}{it}
      \addtoversion{bold}{\mathit}{cmr}{bx}{it}
    \newmathalphabet{\mathcal}
      \addtoversion{normal}{\mathcal}{cmsy}{m}{n}
    \else
    \ifnfsstwo
    \fi
  \fi
\fi

\def\hexnumber#1{\ifcase#1 0\or1\or2\or3\or4\or5\or6\or7\or8\or9\or
 A\or B\or C\or D\or E\or F\fi }

\makeatletter
\ifx\CUP@mtlplain@loaded\undefined
\else
\fi
\makeatother
\makeatletter
\ifx\CUP@mtlplain@loaded\undefined
\else
\fi
\makeatother
\title[H$_2$ in Galaxies]{H$_2$ in Galaxies }

\author[F. Combes]{Fran\c{c}oise  COMBES}

\affiliation{DEMIRM, Observatoire de Paris, 61 Av. de l'Observatoire,
F-75 014, Paris, France}

\begin{document}
\ifnfssone
\else
  \ifnfsstwo
  \else
    \ifoldfss
      \let\mathcal\cal
      \let\mathrm\rm
      \let\mathsf\sf
    \fi
  \fi
\fi

\maketitle

\begin{abstract}
The bulk of the molecular gas in spiral galaxies is under the form
of cold H$_2$, that does not radiate and is only suspected 
through tracer molecules, such as CO. All tracers are
biased, and in particular H$_2$ could be highly underestimated 
in low metallicity regions. Our knowledge is reviewed
of the H$_2$ content of galaxies, according to their types,
environment, or star-forming activities. The HI and CO
components are generally well-mixed (spiral arms, vertical
distribution), although their radial distributions are radically
different, certainly due to radial abundance gradients.
The hypothesis of H$_2$ as dark matter is discussed,
as well as the implications on galaxy dynamics, or the
best perspectives for observational tests.
\end{abstract}

\firstsection 
\section{How to observe H$_2$ in galaxies?}

The bulk of molecular hydrogen in a galaxy is cold, around 10-20K,
 and therefore  invisible. The first rotational level, accessible only through 
a quadrupolar transition, is more than 500 K above the fundamental.
The presence of H$_2$ is inferred essentially from the CO tracer.
The carbon monoxyde is the most abundant molecule after H$_2$; its
dipole moment is small (0.1 Debye) and therefore CO is easily excited,
the emission of CO(1--0) at 2.6mm (first level at 5.52K) 
is ubiquitous in the Galaxy.

\subsection{The H$_2$/CO conversion ratio}

To calibrate the H$_2$/CO ratio, the most direct and natural is
to compare the UV absorption lines of CO and H$_2$ along the same line 
of sight (Copernicus, e.g. Spitzer \& Jenkins 1975;
 ORFEUS, cf Richter et al., this conference). However, only very low
column densities are accessible, in order to see the background source, 
and therefore these observations sample only the diffuse gas, which is
not representative of the global molecular component. It is well
known now that the conversion ratio might vary considerably
from diffuse to dense gas (see below).
The CO molecule is excited by H$_2$ collisions, and should be a
good tracer; but its main rotational lines are most of the times 
optically thick. One can then think of observing its isotopic
substitutes $^{13}$CO or C$^{18}$O, but these are poor tracers
since they are selectively photodissociated, and trace only the
dense cores.

The main justification to use an H$_2$/CO conversion ratio is
the Virial hypothesis: in fact, the CO profiles do not yield the column
densities, but they give the velocity width $\Delta$ V of molecular clouds. 
Once the latter are mapped, and their size R known, the virial mass can
be derived, proportional to $\Delta$ V$^2$ R. There exists a good relation
between the CO luminosity and the virial mass, as shown in Figure 1.
The relation has a power-law shape, but with a slope different from 1.
Both are not proportional, and the conversion ratio should vary by more
than a factor 10 from small to Giant Molecular Clouds (GMC).
In external galaxies, the observations provide only an average over
many clouds, and it has been hoped that the clouds are of the same
nature from galaxy to galaxy. If T$_b$ is the brightness temperature
of the average cloud, the conversion ratio X should vary as n$^{1/2}$/T$_b$,
 where n is the average density of the cloud. This does not take into 
account the influence of the gas metallicity. And the CO luminosity varies
with the metallicity Z, sometimes more than linearly. In the 
Magellanic Clouds, LMC or SMC (Rubio et al 1993), the conversion
ratio X might be 10 times higher than the "standard" ratio. The ratio
can be known for local group galaxies, since individual clouds can
be resolved, and virial masses computed (Wilson 1995).

\subsection{Dust as a tracer}

At millimetric wavelengths, in the Rayleigh-Jeans domain, 
dust emission depends linearly on temperature, and its great advantage is its optical thinness.
In some galaxies, CO and dust emission fall similarly with radius, like
in NGC 891 (Gu\'elin et al. 1993). In other, such as NGC 4565 (Neininger
et al. 1996), the dust emission falls more slowly than CO, although more
rapidly than HI emission. This can be interpreted by the exponential decrease 
of metallicity with radius. The dust/HI ratio follows this dependency, while 
CO/HI is decreasing more rapidly (either due to metallicity, or excitation 
problems).

\subsection{Gamma-rays}

Their emission is proportional to the product of the Cosmic Ray density
and the gas density. But both densities, and their radial profiles are
 not known independently. The $\gamma$-ray radial distribution
is however much more extended than that of the supernovae remnants,
the source of cosmic ray acceleration (e.g. Bloemen 1989).
Recently: EGRET onboard GRO has observed an excess of gamma-rays in the
halo of our Galaxy (see below).

\subsection{Direct H$_2$ observations}

Of course, H$_2$ can also be observed directly
when it is warm. Starbursts and mergers reveal strong 
2.2 $\mu$m emission, like in NGC 6240 (DePoy et al 1986).
The source of excitation has long been debated (X-ray heating, 
UV fluorescence, shocks...) and it was recently concluded that
global shocks were responsible 
(van der Werf et al. 1993, Sugai et al. 1997).
Pure rotational lines have been observed with ISO. In
Arp220, as much as 10\% of the ISM could be in the
warm phase, i.e. 3 10$^9$ M$_\odot$ (Sturm et al. 1996)
while CO observations conclude to a total M(H$_2$) = 3.5 10$^{10}$  
M$_\odot$ (Scoville et al. 1991). In normal galaxies, the warm H$_2$ could be
 less abundant (Valentijn et al. 1996). At least, the 
warm CO component does not affect the H$_2$/CO ratio.

\section{CO and H$_2$ content of galaxies}

From a CO survey of more than 300 galaxies, it has
been concluded that the average molecular content was
comparable to the atomic content:  M(H$_2$)/M(HI)  $\sim$ 1 
(Young \& Knezek 1989; Young \& Scoville, 1991).
But most of galaxies in this survey were selected from their
IRAS flux, and this could introduce a bias.
A recent survey by Casoli et al (1998) near 
the Coma cluster has shown an average
M(H$_2$)/M(HI)  $\sim$ 0.2. 

\subsection{Variation with morphological type}

It is well established that the HI component is proportionally
more abundant relative to the total mass in late-type
galaxies. The opposite trend is observed for the H$_2$,
at least as traced by the CO emission.
M(H$_2$)/M(HI) is therefore smaller for late-types,
by a factor $\sim$ 10. However, this 
could be entirely a metallicity effect. Since the 
metallicity is increasing with the mass of the galaxy,
a test is to select the most massive galaxies of
late-type. For these high-mass galaxies, there is no trend of decreasing 
H$_2$ fraction with type (Casoli et al. 1998).

\begin{figure} 
\centerline{\psfig{figure=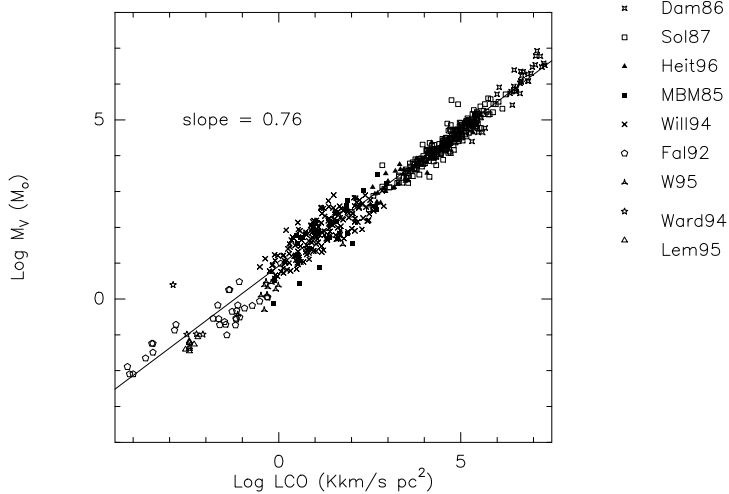,bbllx=25mm,bblly=1cm,bburx=165mm,bbury=22cm,width=10cm,angle=-90}}    
  \caption{ Virial mass versus CO luminosity for molecular clouds in the Milky Way.
The fit corresponds to M$_{\rm V} \propto$ L$_{\rm CO}^{0.76}$. The data are from
Dam86:  Dame et al. (1986); Sol87: Solomon et al. (1987);
Heit96: Heithausen  (1996); MBM85: Magnani et al. (1985);
Will94: Williams et al. (1994); Fal92: Falgarone et al. (1992);
W95: Wang et al. (1995); Ward94: Ward-Thompson et al. (1994);
Lem95: Lemme et al. (1995).}
\end{figure} 

\subsection{ Dwarf and LSB galaxies}

The strong dependency of the H$_2$/CO conversion ratio on
metallicity Z is also the main problem in the observations
of dwarf and Low Surface Brightness (LSB) galaxies. Both
have low metallicity. It appears that the 
conversion factor X can vary linearly and even more
with metallicity, as predicted by Maloney \& Black (1988). 
Not only, the low abundance of C and O lowers the
abundance of CO, but also the dust is less abundant,
and therefore the UV light is less absorbed, and
spread all over the galaxy, photo-dissociating the CO molecules. 
When the dust is depleted by a factor 20, there should be only
 10\% less H$_2$, but 95\% less CO (Maloney \& Black 1988).

In dwarf galaxies, CO emission is very low,
and it is difficult to know the H$_2$ content.
If the HI/H$_2$ ratio is assumed constant from
galaxy to galaxy, then 
X varies with Z$^{-2.2}$ (Arnault et al 1988).
Recent results by Barone et al (1998),
Gondhalekar et al (1998) and Taylor et al (1998)
confirm this strong dependency on metallicity,
increasing sharply below 1/10th of solar metallicity.

Low-surface-brightness galaxies have
large characteristic radii, large gas fraction
and are in general dark matter dominated;
they are quite unevolved objects.
Their total gas content is similar to that of
normal galaxies (McGaugh \& de Blok 1997).
But CO is not detected in LSB (de Blok \& van der Hulst 1998).
Due to their low surface density, below the threshold
for star formation, these galaxies have a very
low efficiency of star formation (Van Zee et al 1997).
The cause could be the absence of companions, since
LSB live in poor environments (Zaritsky \& Lorrimer 1993).
It is well known that galaxy interactions, by driving in 
a high amount of gas, trigger star formation.

\subsection{Ultra-luminous IRAS galaxies} 

At the opposite, there exists a class of galaxies, characterized
by their bursts of star formation; these are 
ultra-luminous in far-infrared, because of the emission of
dust heated by the new stars. These objects possess
large amounts of gas, particularly condensed in
the inner parts, certainly due to  interactions and mergers.
CO emission is highly enhanced in these starbursting galaxies, 
and large H$_2$ masses are deduced, even with a modified
(lower than standard) conversion ratio (Solomon et al 1997).
The prototype of these objects is the nearby Arp220: 
new CO interferometer data show that CO is in rotating nuclear 
disks (Downes \& Solomon 1998), where the surface density 
of gas is about 30\% of the total surface density.

\section{Spatial distribution}

\subsection{Radial Distributions}

The differences between HI and H$_2$ (or CO) radial
distributions in galaxies is striking (cf figure 2).
While all components related to star formation,
the blue luminosity from stars, the H$\alpha$ (gas ionised
by young stars), the radio-continuum (synchrotron related
to supernovae), and even the CO distribution, follow
an exponential distribution, the HI gas alone is
extending much beyond the ``optical'' disk, sometines
in average by a factor 2 to 4 (R$_{HI}$ = 2-4 R$_{opt}$).
The HI gas has very often a small deficiency in the center.
 Would this mean that the atomic gas is transformed in
molecular phase in the denser central parts? This is 
possible in some galaxies, where the HI and CO distribution
appear complementary, but it is not the general case, 
all possibilities have been observed, including a central 
gaseous hole, both in CO and HI (like in M31, for example).

\begin{figure} 
\centerline{\psfig{figure=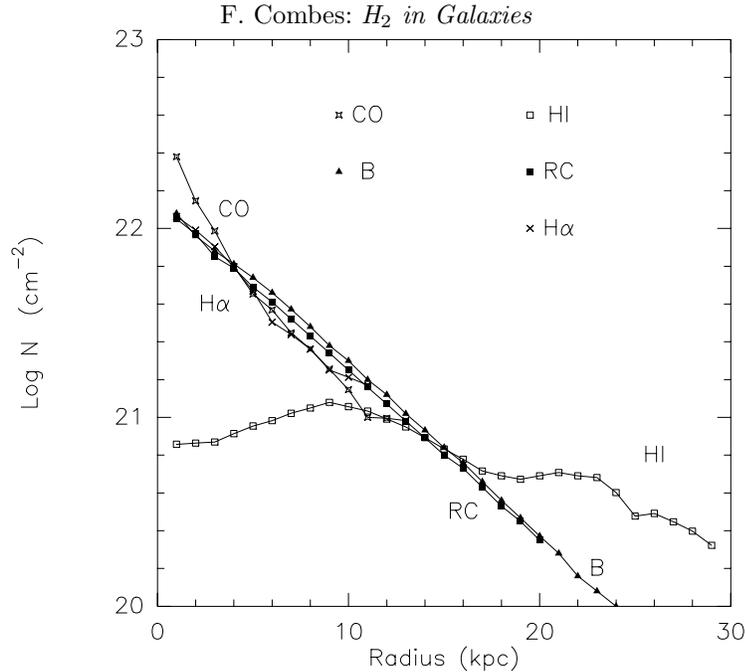,bbllx=25mm,bblly=1cm,bburx=165mm,bbury=17cm,width=10cm,angle=-90}}    
  \caption{ Radial distributions of various surface densities in
a typical spiral galaxy NGC 6946: H$_2$(CO) and HI column densities,
Blue, Radio-continuum and H$\alpha$ surface densities
(adapted from Tacconi \& Young, 1986).}
\end{figure}

\subsection{Large-scale Structure}

Within the optical disk, where CO is observed easily,
there is a very good large-scale correlation between both 
gas components (see e.g. Neininger et al. 1998). They appear
well mixed, and follow the spiral ams with large contrast.
This is also true for the ionised gas (HII regions).
Of course, this is only at 100pc-1kpc scale; at very small scale
the various components can be anti-correlated, the HI
gas being found more at the envelopes of dense molecular
clouds, the ionised gas being also anti-correlated with
the neutral gas.

\subsection{Vertical Structure}

In our own Galaxy, and in external galaxies seen edge-on,
the galaxy disks appear much narrower in CO emission
than in HI. This suggests that the molecular gas is more
confined to the plane, and that its vertical dynamical oscillations
are of less amplitude than for the atomic gas. The consequence
should be a vertical velocity dispersion much lower for 
the molecular gas, since for a given restoring force due to the 
stellar disk, the maximum height above the plane is 
proportional to the z-velocity dispersion. Surprisingly,
this is not the case: in face-on galaxies both CO 
(Combes \& Becquaert 1997) and HI (Kamphuis 1992) velocity dispersions
are observed of similar values ($\sigma_v \sim 6$ km/s),
and remarkably constant with radius. This is not a saturation 
effect of the CO lines, since the $^{13}$CO spectra show the same.
A possible interpretation is that both gas are well mixed,
in fact it is the same dynamical component, which changes
phase along its vertical oscillations. It is possible that
the H$_2$ gas follows the HI, but the CO is photo-dissociated
at high altitudes, or not excited. Or even the H$_2$ could
disappear, since the chemistry time-scale  ($\sim$ 10$^5$ yr)
is much smaller than the dynamical z-time-scale
($\sim$ 10$^8$ yr).

\subsection{Small scale structure of clouds}

The molecular component is also characterized by
its remarkable self-similar structure, a hierarchical 
system of clouds, tightly related to a fractal structure.
It can be quantified by power-law relations between
cloud size and linewidth, or size and mass (Larson, 1981).
These relations are observed, whatever the radial distance 
to the center, and in the HI component as well.
The fact that the same structure is observed 
outside of the star-forming regions is puzzling:
the HI gas outside the optical disk displays a very clumpy structure,
implying that it is unstable at all scales (spiral arms, self-similar
structure of clumps). The fact that these gravitational instabilities
do not trigger star formation must be explained.

\section{H$_2$ as a dark matter candidate}

One of the driver to propose cold H$_2$ as a dark matter
candidate is our increasing knowledge about 
evolution of galaxies along the Hubble sequence 
(e.g. Pfenniger, Combes \& Martinet, 1994). Because of
spiral waves and bars, galaxies progressively concentrate their
mass towards the center, and the late-type galaxies evolve
to early-types in the sequence. Besides, HI observations of rotation
curves have shown that the fraction of dark matter in the total mass
is larger in late-type galaxies: therefore, some of the dark matter
must be transformed into stars during evolution (cf
Pfenniger, this conference).

\subsection {Baryonic mass fraction }

The quantity of baryons in the Universe (and more
precisely the fraction of the critical density in baryons 
$\Omega_b$) is constrained by the primordial nucleosynthesis
to be $\Omega_b =$ 0.013 h$^{-2}$, with h = H$_0$/(100 km/s/Mpc)
is the reduced Hubble constant. With h = 0.5, $\Omega_b$
is 0.09, and more generally $\Omega_b$ is between 0.01 and 0.09
(Walker et al. 1991, Smith et al. 1993), while the 
visible matter corresponds to  $\Omega_* \sim $0.003 (M/L/5) 
h$^{-1}$ (+ 0.006 h$^{-1.5}$ for hot gas). Therefore, most
of the baryons (90\%)  are dark.
 
In rich clusters of galaxies, the baryons are more visible, 
under the form of hot gas, they constitute $\sim$ 30\% of the total mass 
(White et al. 1993). Since clusters must be representative
of the baryonic fraction of the Universe, this implies that the total 
mass cannot be larger than 3 times the baryons mass 
(or $\Omega_m < 0.3$).

\subsection{ The smallest fragments }

The existence of a large number of gas clumpuscules (of $\sim$ 10 
AU in size) in the Galaxy has already been invoked to 
explain the observed ESE (Extreme Scattering Events) in front of
quasars, by Fiedler et al. (1987, 1994). About 
300 QSOs were observed during a few years, 150 over 12 yrs.  
More than 10 ESE events were detected, due to diffraction or 
refraction by a region of high electronic density (n$_e$).
From the duration of the events, sizes of $\sim$ 10 AU are derived,
and from their frequency, the number of clumpuscules in the Galaxy
must be about 10$^3$ the number of stars. The neutral density
of these objects is still a matter of debate. Their stability is
best explained in the hypothesis that they are self-gravitating.
The mass of one clumpuscule is then of the order of 10$^{-3}$ M$_\odot$.
Walker \& Wardle (1998) have recently built models of self-gravitating clouds, 
with envelopes ionised by the interstellar radiation field: they 
found for the electronic density the right order of magnitude
to account for the observed ESE. This hypothesis is supported by
direct observations through HI VLBI in absorption  in front of
remote radio sources (Diamond 
et al. 1989, Davis et al. 1998, Faison et al. 1998); large column
densities ($\sim 10^{21}$ cm$^{-2}$) are observed with
sizes of $\sim$ 10 AU, leading to HI densities of 10$^6$ cm$^{-3}$ or more. 

\subsection{ Gamma-rays }

Dixon et al. (1998) from EGRET observations have 
recently detected an excess of diffuse $\gamma$-ray emission 
in the galactic halo. This could be interpreted in several ways:
either coming from un-resolved sources associated to the Galaxy;
or being due to high-latitude inverse compton emission; or finally to extra
molecular gas in the halo, through cosmic ray/nucleon reaction
giving $\pi_0$ then $\gamma$-rays. This has been developped
by de Paolis et al. (1999) and Kalberla et al. (1999), see also
Shchekinov (this conference). Cosmic rays are stopped by thick
clumpuscules, that have enough column density
to be opaque for both cosmic rays and gamma rays.
Sciama (1999) proposes that cosmic rays 
are fragmented in clouds, heat the clouds, and are responsible 
for the their FIR emission (Sciama 1999). However,
the absorbed energy is non negligeable, and since clouds in 
these halo models are assumed to move through the optical plane
(in their z-ocillations), sweep up high-metallicity gas, and
therefore contain CO molecules, they should be visible
through CO emission.

\subsection{Various models of H$_2$ as dark matter}

The first model proposes to prolonge the visible gaseous disk 
towards large radii, with thickening and flaring, following the HI disk.
 The cold and dark H$_2$ component is supported by rotation,
exists only outside the optical disk, where it is required by rotation curves
(Pfenniger et al 1994, Pfenniger \& Combes 1994). The gas is stabilised
through a constantly evolving fractal structure, experiencing
Jeans instabilities at all scales, in thermal equilibrium with the 
cosmic background radiation at T = 2.7 (1+z) K.

Other models distribute the dark molecular gas in a spherical or
 flattened halo, with no hole within the optical disk. The molecular
gas is not so cold, and is associated with clusters of brown dwarfs
or MACHOS (de Paolis et al. 1995, Gerhard \& Silk 1996, Shchekinov,
this conference).

In the clumpuscule model, the HI gas can be considered as a tracer, 
the interface between the molecular clumps and the extra-galactic radiation field.
Beyond the HI disk, there could be an ionization front, and the interface might
become ionized hydrogen. In this context, there should exist a distribution 
correlation between the dark matter and the HI gas. This is indeed the
case, as already remarked by Bosma (1981), Broeils (1992) or Freeman (1993):
there is a constant ratio between the surface density of dark matter, as
deduced from the rotation curves, and the HI surface density,
$\Sigma_{DM} / \Sigma_{HI}$ =  7-10 (cf figure 3, and a recent work
by Hoekstra et al. 1999). This
ratio is constant with radius in a given galaxy, and varies 
slightly from galaxy to galaxy, being larger in early-types.
However, the dark matter does not dominate the mass in the latter, and
therefore the estimate of its contribution is more uncertain. 
The correlation is the most striking in dwarf galaxies, which are
dominated by dark matter. The observed velocity curve is
almost exactly proportional to the velocity curve expected from the HI 
component alone. Figure 3 shows the example of NGC 1560, 
from Broeils (1992). 
 Let us note that dwarf galaxies represent a hard test for all models of
dark matter, since the stellar component does not dominate the mass.
They rule out cold dark matter (CDM) profiles (Burkert \& Silk 1997), and 
hot dark matter (HDM) models are also unable to concentrate as much as is observed
(Lake 1989, 1990, Moore 1996). Baryonic dark matter is thus required.

\begin{figure} 
\centerline{\psfig{figure=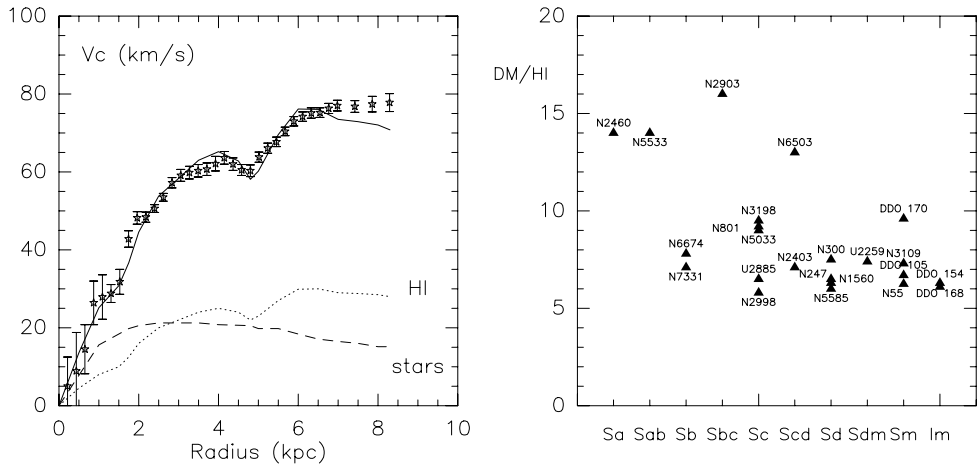,bbllx=5cm,bblly=1cm,bburx=16cm,bbury=24cm,width=13cm,angle=-90}}    
  \caption{ {\it Left}: HI rotation curve of NGC 1560 (dots +error-bars), with the
rotation curve due to the HI itself (dotted line) and the stellar component (dash).
The full line is the resulting expected rotation curve, when the HI mass has been
multiplied by 6.2. {\it Right}: The ratio of surface densities of dark matter to HI
required to explain the rotation curve of galaxies, as a function of type. Data from 23
galaxies have been taken from Broeils (1992) and references therein.}
\end{figure} 

\subsection{Detection possibilities}

If there exists a transition region where the cold H$_2$ is
mixed in part with evolved gas with enough metallicity and
dust, it might be possible to detect cold dust emission.
Encouraging results have been found by the COBE satellite,
concluding to the existence of a cold (4-7K) component with 
column densities 10 times that of the warm component
(at 18K), and more confined to the outer Galaxy
(Reach et al. 1998).

Another promising tool for detection is H$_2$ absorption in 
the UV electronic lines.
The main problem is the low expected surface filling factor 
of the cold gas ($\sim$ 1\%).
H$_2$ has already been detected in front of QSO 
through intervening galaxies (Foltz et al. 1988, Ge et al 1997); but
these observations suffer from severe confusion problem in the 
Ly$\alpha$ forest. With the Hubble Space telescope, it was not possible
to observe the fundamental lines at zero redshift, but it will be possible
with FUSE. Let us remark that heavy lines of sight will be impossible
to observe, due to obscuration of the background source
(e.g. Combes \& Pfenniger 1997). Finally, observations of
the lowest pure rotational lines of H$_2$ have suggested some clues
for the existence of large quantities of H$_2$ in galaxies
(Valentijn \& van der Werf 1999) and should be pursued
in external galaxies, at much further radius than was possible with ISO.

\section{Conclusions} 
 
The bulk of the H$_2$ mass in galaxies is cold and furtive.
The main tracer is the CO molecule, but the H$_2$/CO conversion 
ratio is very variable, according to the physical conditions in molecular clouds
(density and temperature), but mainly with the metallicity.

Probably because of radial abundance
gradients, the molecular gas traced by the CO emission 
is only observed to extend over the optical disk,
while the atomic gas component is prolonged much farther
 out in radius. Nevertheless, the 
HI and H$_2$ at the same radius are tightly correlated,
at large-scale (kpc scale). They trace the same spiral
structure for example. In the vertical direction, the two components
are also well mixed: both reveal a constant vertical velocity dispersion
with radius, of comparable amplitude.
This suggests that the HI could serve as 
the tracer of the H$_2$ component,
that would then also extends far out in radius. The constant
ratio between dark matter and HI surface densities in galaxies
support this hypothesis.
If cold molecular gas is a good candidate of baryonic dark
matter, observational tests should be pursued: H$_2$ UV
absorption lines may be the best probe, and data from 
the FUSE satellite will certainly make big advances
on the subject.


\end{document}